\documentclass[a4paper,fleqn]{mnras}
\usepackage{amsmath,amsfonts,amsopn,amssymb,stmaryrd,xspace}
\usepackage{tablefootnote}
\usepackage{txfonts}
\usepackage[T1]{fontenc}
\usepackage{ae,aecompl}

\usepackage{graphicx}	% Including figure files
\usepackage{amssymb}	% Extra maths symbols

\title[CHARA resolves WR 137 and WR 138]{The CHARA Array resolves the long-period Wolf-Rayet\\ binaries WR 137 and WR 138}

\author[N. D. Richardson et al.]{Noel~D.~Richardson$^{1}$\thanks{E-mail:noel.richardson@utoledo.edu},
Tomer Shenar$^2$,
Olivier Roy-Loubier$^3$,
\newauthor Gail Schaefer$^{4}$,
Anthony F. J. Moffat$^{3}$,
Nicole St-Louis$^{3}$,
Douglas R. Gies$^{5}$, 
\newauthor Chris Farrington$^{4}$,
Grant M. Hill$^{6}$, 
Peredur M. Williams$^{7}$,
Kathryn Gordon$^{5}$, 
\newauthor Herbert Pablo$^3$, and
Tahina Ramiaramanantsoa$^3$ \\ 
$^{1}$ Ritter Observatory, Department of Physics and Astronomy, The University of Toledo, Toledo, OH 43606-3390, USA\\
$^{2}$ Institut f\"{u}r Physik und Astronomie, Universit\"{a}t Potsdam, Karl-Liebknecht-Str. 24/25, D-14476 Potsdam, Germany\\
$^3$ D\'epartement de physique and Centre de Recherche en Astrophysique du Qu\'ebec (CRAQ), Universit\'e de Montr\'eal, C.P. 6128,\\  Succ.~Centre-Ville, Montr\'eal, Qu\'ebec, H3C 3J7, Canada\\
$^{4}$ The CHARA Array, Mount Wilson Observatory, 91023 Mount Wilson CA, USA \\
$^{5}$ Center for High Angular Resolution Astronomy, Department of Physics and Astronomy, Georgia State University, P. O. Box\\ 5060,  Atlanta, GA 30302-5060, USA \\
$^{6}$ W. M. Keck Observatory, 65-1120 Mamalahoa Highway, Kamuela, HI 96743, USA\\
$^{7}$ Institute for Astronomy, University of Edinburgh, Royal Observatory, Edinburgh EH9 3HJ, UK \\
}

\begin{document}
%%\received{} 
%%\accepted{} 
\bibliographystyle{mn2e}

\date{}

\pagerange{\pageref{firstpage}--\pageref{lastpage}} \pubyear{2016}

\maketitle

\label{firstpage}

\begin{abstract}

We report on interferometric observations with the CHARA Array of two classical Wolf-Rayet stars in suspected binary systems, namely WR 137 and WR 138. In both cases, we resolve the component stars to be separated by a few milliarcseconds. The data were collected in the $H$-band, and provide a measure of the fractional flux for both stars in each system. We find that the WR star is the dominant $H$-band light source in both systems ($f_{\rm WR, 137} = 0.59\pm0.04$; $f_{\rm WR, 138} = 0.67\pm0.01$), which is confirmed through both comparisons with estimated fundamental parameters for WR stars and O dwarfs, as well as through spectral modeling of each system. Our spectral modeling also provides fundamental parameters for the stars and winds in these systems. The results on WR 138 provide evidence that it is a binary system which may have gone through a previous mass-transfer episode to create the WR star. The separation and position of the stars in the WR 137 system together with previous results from the IOTA interferometer provides evidence that the binary is seen nearly edge-on. The possible edge-on orbit of WR 137 aligns well with the dust production site imaged by the {\it Hubble Space Telescope} during a previous periastron passage, showing that the dust production may be concentrated in the orbital plane.   

\end{abstract}


\section{Introduction}

Every star is described with a set of fundamental parameters including its radius, $R$, mass, $M$, luminosity, $L$, and effective temperature, $T_{\rm eff}$. For most stars, we must estimate these from accurate photometry, a measure of the distance, and a reliable spectrum. This can be calibrated using techniques involving binary stars where we either utilise an eclipsing system to understand fundamental parameters or where a visually resolved orbit can allow us to infer the stellar masses of the binary. Recent progress has been made in the area of fundamental parameters through interferometric techniques (e.g., Boyajian et al.~2012) where long-baseline interferometry can resolve single stars, yielding direct measures of angular diameters. When coupled with an SED and distance, measurements of stellar radii and temperatures are possible to accuracies of a few percent allowing accurate predictions for other stars (Boyajian et al.~2014). 

Empirical relations for massive stars are more complicated to determine due to their scarcity and large distances. Even with a large binary fraction (e.g., Aldoretta et al.~2015, Sana et al.~2014), the calibration of stellar parameters has relied on eclipsing binaries. Sana et al.~(2012) show that most massive stars with periods of $P\lesssim 1000$ d will either interact in the future or have already interacted. Therefore, most stars that are used to calibrate the mass-radius-luminosity relationships for massive stars (Martins et al.~2005) are not representative of single-star evolution. If we wish to understand the evolution of massive stars without binary effects, we need to measure directly masses for evolved stars in long-period, widely-separated binaries. In order to measure the masses of these stars, as well as place lower limits on the mass loss of the more-evolved component, we must visually resolve the orbits of nearby systems where we can also obtain a double-lined spectroscopic orbit.

The Wolf-Rayet (WR) stars represent a class of objects with great potential to determine the empirical lower limits to the mass lost by massive stars, which Smith \& Owocki~(2006) speculate could be up to 65\% of the mass of a 60 $M_\odot$ star prior to the supernova explosion. Thus far, only two WR systems have been examined with interferometry. They are $\gamma^2$ Velorum (Hanbury Brown et al.~1970, Millour et al.~2007, North et al.~2007) and WR 140 (Monnier et al.~2004, 2011). Both systems have well-defined double-lined spectroscopic orbits (Schmutz et al.~1997, Fahed et al.~2011) and consist of a carbon-rich WC star orbiting an O star. For $\gamma^2$ Vel, the spectral types of the component stars are WC8$+$O7.5III (Schmutz et al.~1997), while for WR 140, they are WC7pd$+$O5.5fc (Fahed et al.~2011).

The first results on $\gamma^2$ Vel (WR 11, HD 68273) were presented by Hanbury Brown et al.~(1970), who found that the Narrabri Intensity Interferometer could be used not only to resolve single hot stars as it was designed for, but also to resolve binary orbits and determine the angular sizes of the emission-line forming regions for some hot stars. As such, they derived a separation of the binary ($P=$78.5 d) of $4.3\pm0.5$ mas, corresponding to a distance of $350\pm50$ pc when compared to the double-lined spectroscopic orbit. Further, Hanbury Brown et al.\ found that the emission line region was 0.24$a$ in relation to the semi-major axis, $a$. $\gamma^2$ Vel was then studied with VLTI/AMBER in the near-infrared by Millour et al. (2007). These results proved that the expected visual orbit was fairly well-constrained at the time of the observation, and that the predicted WR and O star fluxes were compatible with the previous modeling of the system (de Marco et al.~1999, 2000). Further, they found that roughly 5\% of the flux in the $K-$band was from the free-free emission in the wind-wind collision region. Finally, North et al.~(2007) observed $\gamma^2$ Vel across its entire orbit with the Sydney University Stellar Interferometer (SUSI) to obtain a visual orbit and measure the distance to an unprecedented precision of $336^{+8}_{-7}$ pc. The resulting masses were $M_{\rm WR}=9.0\pm0.6 M_\odot$ and $M_{\rm O}=28.5\pm1.1 M_\odot$, and represent the best-measured mass for any WR star.

WR 140 (HD 193793) is a prototype for colliding wind systems with a highly elliptical ($e=0.896$), long period ($P=7.93$ years) orbit. The system was resolved with the IOTA3 Imaging Interferometer by Monnier et al.~(2004), and then Monnier et al.~(2011) combined the IOTA3 measurements and new CHARA measurements with the radial velocity orbit based on an intensive spectroscopic campaign by Fahed et al.~(2011) to measure the visual orbit precisely and then derive the masses of the component stars to be $M_{\rm WR}=14.9\pm0.5 M_\odot$ and $M_{\rm O}=35.9\pm1.3 M_\odot$, with the system at a distance of $1.67\pm0.03$ kpc.

Two long-period WR stars are obvious candidates for follow-up long-baseline interferometry: WR 137 and WR 138.
WR 137 (HD 192764) consists of a WC7pd + O9 binary with a period of 13.05 years and a small eccentricity of  $e =  0.18$ (Lef\`evre et al.~2005). It has a known RV orbit (at least for the WR component; the O star's orbit is very noisy) with $a\sin i = 15$ AU, which combined with a distance of 1.82 kpc (Nugis \& Lamers 2000), yields a value of $\sim$8 mas, well within the reach of modern interferometers. In fact, a preliminary interferometric result presented by Rajagopal (2010) indicates that the binary was resolved by the IOTA interferometer in the $H-$band, with a flux ratio measured relative between the two stars of $f_2/f_1$=0.81 and a separation of 9.8 mas in 2005, although the flux ratio is ambiguous as to which component star is represented by which. This makes this system appealing to observe with follow-up observations so that a third WR+O system can be resolved. Further, the system is a known dust producer (Williams et al.~1985, Marchenko et al.~1999, Williams et al.~2001), so multi-wavelength interferometric observations could reveal the location of dust formation in these particular colliding wind systems that form dust spirals (e.g., Tuthill et al.~2008).

WR 138 (HD 193077) is a potential long-period binary consisting of a WN5o (Smith et al.~1996) and an O9 star (Annuk 1990), although the classification of the secondary has not been well constrained. Annuk (1990) presented the orbital elements for the Wolf-Rayet and O components. The putative orbit has a period of $P = 4.21$ years (1538 d) and $e = 0.3$. The period was confirmed to be 1521$\pm$35 d with additional data presented by Palate et al.~(2013), although their newer data were too sparse to better fit the orbital elements. Further, Palate et al.~(2013) found that the X-ray emission was fully consistent with a colliding winds binary. The Wolf-Rayet radial velocity curve is much better defined than that of the O star, which shows an apparent, low-amplitude, variation in anti-phase to the WR star. With a measurement of $a \sin i = 4.3$ AU, we expect a separation of $\sim$2.2 mas when combined with an estimated distance of 1.9 kpc (from the mean of that for WR 133 and WR 139 in the same association\footnote{as listed in Rosslowe \& Crowther (2015) at http://pacrowther.staff.shef.ac.uk/WRcat/}). 
However, the O-component absorption lines of WR 138 are very broad and shallow, leading Massey (1980) to claim that WR 138 was a single, newly formed WR star with intrinsic broad absorption lines. Interferometry can resolve the issue by either resolving two stars or showing an unresolved star. With very broad absorption lines, the secondary would more likely be a main sequence star. 

In this paper, we present interferometric observations of WR 137 and WR 138 with the CHARA Array that show that both of these systems are resolved with long-baseline near-infrared interferometry. Section 2 outlines our interferometric observations. In Section 3, we show the binary fits for the observations. We discuss these systems in Section 4, and present goals of future studies and conclude this paper in Section 5.

\section{Long Baseline Interferometry with the CHARA Array}

We collected long baseline near-infrared interferometry of WR 137 and WR 138 using the CHARA Array (ten Brummelaar et al. 2005) and CLIMB beam combiner (ten Brummelaar et al.~2013) in the $H-$band during the nights of 2013 August 13--14. The CHARA Array is a $\Ydown$-shaped interferometric array of six 1-m telescopes with baselines ranging from 34 to 331 meters in length. Our observations used longer baselines consisting of the S1, E2, and W1 telescopes which yield maximum baselines between 251 and 278 meters. The observations are summarized in Table 1.

\begin{figure}
\includegraphics[width=2.5in, angle=90]{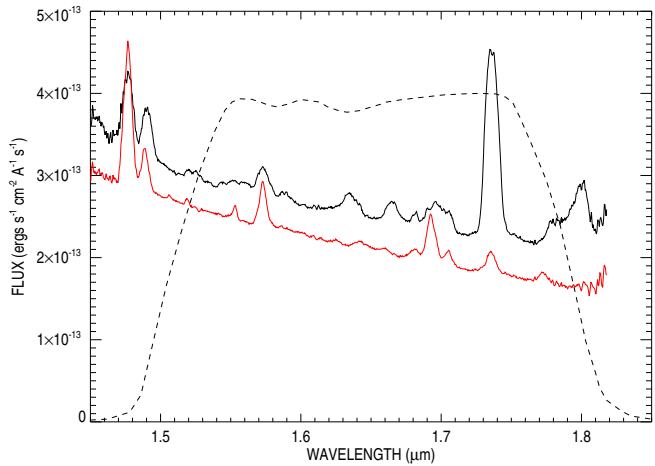}

\caption{\label{nirspec} NIR spectra of WR 137 (black) and WR 138 (red) from Shara et al.~(2012) are shown, with an $H$-band filter response function overplotted as a dashed line. For WR 137, the strong emission line of C\,{\sc iv}\ $\lambda 1.736\mu$m can drastically alter the instrumental response of the CHARA Array, and were accounted for through comparison of the Fourier spectrum of the fringe envelopes of the calibrator and target stars. In contrast, the weaker lines in the spectrum of WR 138 account for very little difference in the instrumental response. }

\end{figure}

\begin{table*}
\centering
\begin{minipage}{180mm}
\caption{CHARA Interferometric Observing Log \label{table2}}
\begin{tabular}{llllll}
\hline %ADD FILTERS!!!!
UT Date		&		Baselines			& Baseline Length(s)	 [m]	&	N$_{137}$		&	$N_{138}$	&	 	Calibrator HD number(s)		\\
\hline
2013 Aug 14	&	 	S1/E2/W1			&	278, 251, 278		&	2				&	2			&		191703, 192804	\\ %HD 192641
2013 Aug 15	&		S1/E2/W1			&	278, 251, 278		&	3				&	0			&		191703, 192804, 192536 	\\ %HD 192641, 193077

\hline
\end{tabular}
\end{minipage}
\end{table*}

\begin{figure*}
\includegraphics[width=4in]{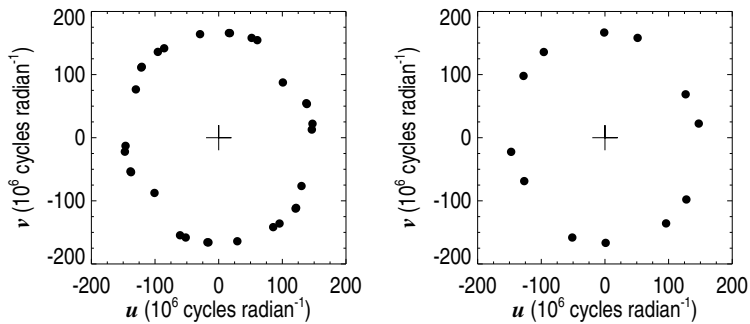}

\caption{\label{uv} The on-sky $(u,v)$ coverage for WR 137 (left) and WR 138 (right). While fewer observations were made on WR 138, the $(u,v)$ plane is still well-covered.  }

\end{figure*}

To measure the instrument response and calibrate our data, we observed calibrator stars with small angular diameters both before and after each observation of a target star. Namely, we observed the calibrator stars listed in Table 2. 
In order to calibrate correctly our data, we fit these stars' spectral energy distributions with Kurucz model atmospheres corresponding to the spectral types previously published. The resulting angular diameters, and published spectral types are also given in Table 2, in good agreement with Lafrasse et al.~(2010), and are all unresolved by the CHARA Array in the $H$-band, where the resolution limit for a single star is $\approx$ 0.5 mas. 

\begin{table*}
\centering
\begin{minipage}{180mm}
\caption{Calibrator Stars \label{table2}}
\begin{tabular}{lllll}
\hline 
HD number	& 		Spectral Type		& Reference						&	Angular Diameter [mas]	& Nights used	\\
\hline
191703		&		F0V			&	Honeycutt \& McCuskey (1966)			&	$0.21\pm0.01$ & 13, 14 Aug 2013 \\
192804		&		F8V			&	Ljunggren \& Oja (1961)				&	$0.23\pm0.01$ & 13, 14 Aug 2013 \\
192536		&		A7III			&	Barbier (1963)						&	$0.17\pm 0.01$	& 14 Aug 2013 \\
\hline
\end{tabular}
\end{minipage}
\end{table*}

The data were reduced using the standard CHARA reduction pipeline (ten Brummelaar et al.\ 2005, 2013).  The visibilities and closure phases were averaged over each observing block. The calibrated OIFITS data files (Pauls et al.~2005) will be available through the JMMC archive\footnote{http://www.jmmc.fr/oidb.htm}. Wolf-Rayet stars have strong emission lines throughout the optical and NIR spectrum (Fig.~1). These emission lines reduce the effective bandpass over which the fringe amplitude is measured causing the true visibility to be smaller than if a fixed bandpass was assumed. We measured the effective bandpass through comparisons of the width of the power spectra of both calibrator stars and the targets, giving a reasonable approximation to the instrumental response. This effect caused the visibilities of WR 137 to change by $-$15\%. The data collected on WR 138 were relatively unaffected by this, and any changes could be accounted for in the measurement errors.

Importantly, the observations are well-sampled on the $(u,v)$ plane (Fig.~2), meaning that any fits to the data should provide meaningful measurements of the separation, position angle, and flux ratio for the two stars. We also note that further measurements of these systems were not possible during the observing window as a target-of-opportunity (Nova Del 2013) became a priority (Schaefer et al.~2014). However, the observations obtained allow for measurements of the binary nature of these stars.

\section{Resolving the Long-Period Binaries WR 137 and WR 138}

\begin{table*}
\centering
\caption{Binary Fits \label{table4}}
\begin{tabular}{lcclcccl}
\hline
Star 		& Separation (mas)	&	PA ($^\circ$)	&	Date			& Phase	&	$f_{\rm WR}$	& $f_{\rm O}$	& Reference \\ \hline
WR 137	&	9.8$\pm$0.6	&	295$\pm$1.3	&	2005 July		&	0.70	&	0.56$\pm$0.2		&	0.44$\pm$0.2		& Rajagopal (2010) \\
WR 137	&	4.03$\pm$0.02	&	131.6$\pm$0.3	&	2013 August	&	0.33 	&	0.59$\pm$0.04		&	0.41$\pm$0.04		& this work \\
WR 138	&	12.37$\pm$0.04	& 305.3$\pm$0.2	&	2013 August	& 0.31 & 	0.67$\pm$0.01		&	0.33$\pm$0.01		& this work	\\ \hline
\end{tabular}
\end{table*}

\begin{figure*}
\includegraphics[width=6.4in, angle=0]{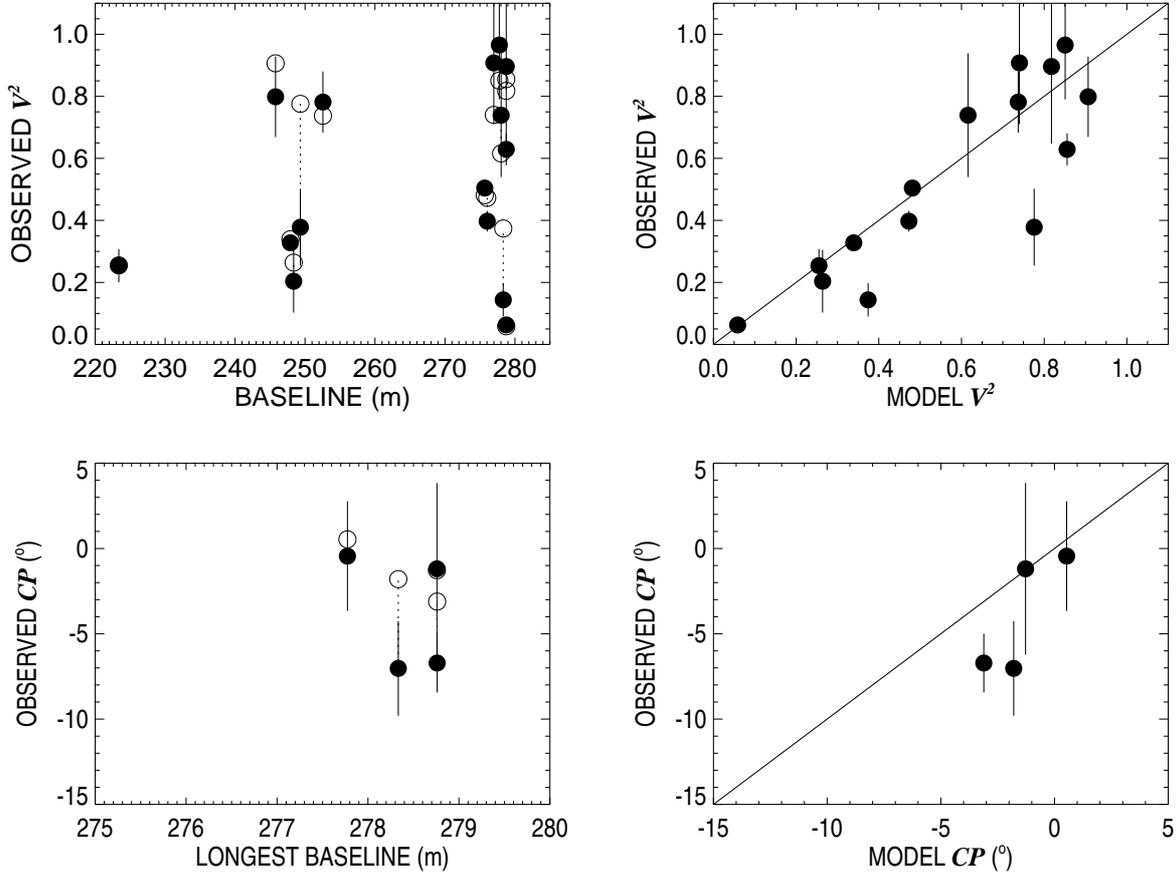}

\caption{\label{wr137} CHARA observations of WR 137. The upper left plot shows the squared visibilities compared with the baseline, where the solid points represent the measurements and the open circles connected by dotted lines represent the best model (Table 3) for the binary fit. The upper right compares the model and the measurements, with a 1:1 relation overplotted. The lower left panel shows our four measurements of closure phase (solid points) compared with the longest baseline from the three telescope configuration, with the model points connected in the same manner as the upper left panel. The lower right panel shows the comparison of the model and the measurements. }
\end{figure*}

\begin{figure*}
\includegraphics[width=6.4in, angle=0]{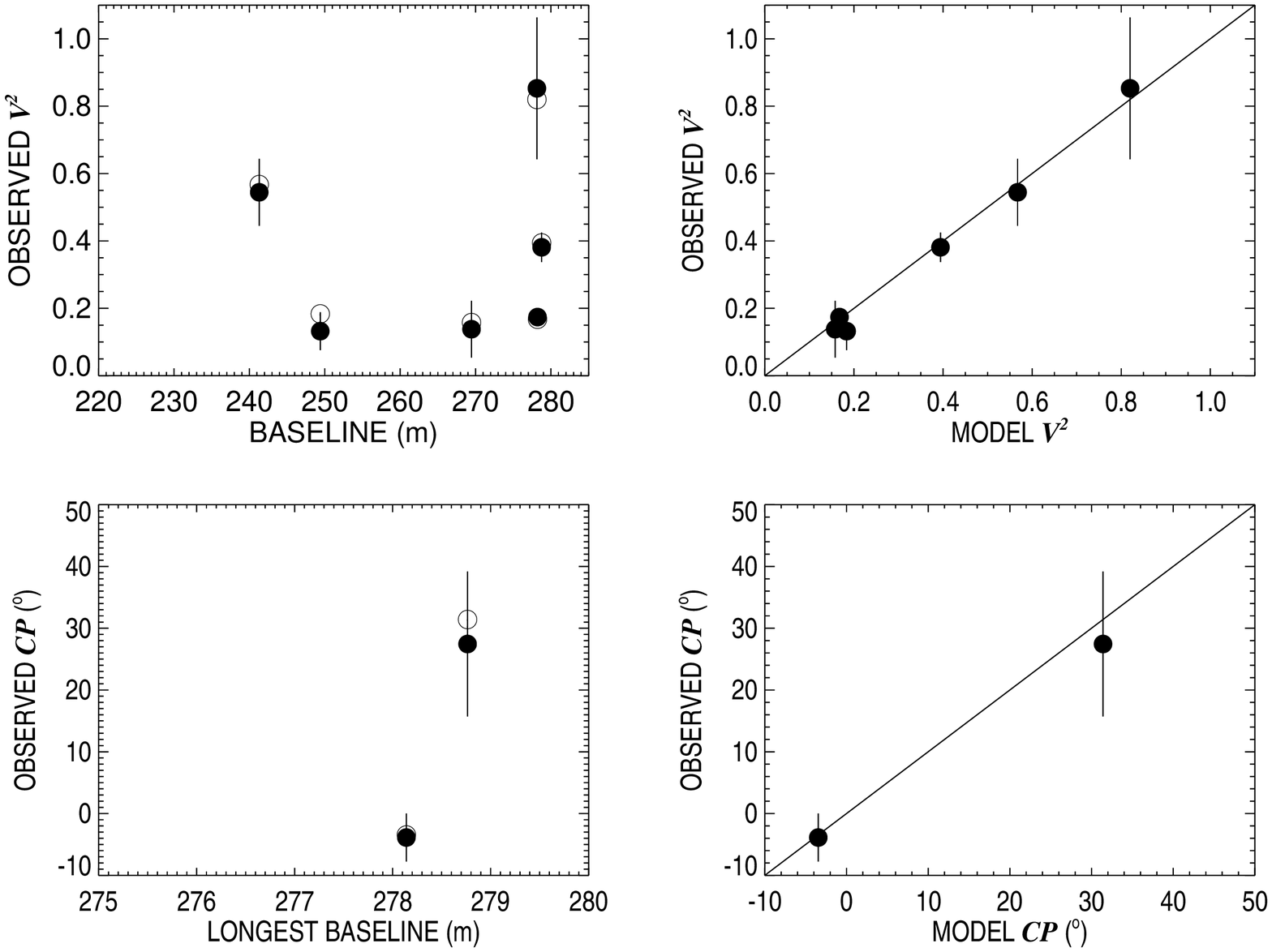}

\caption{\label{wr138} CHARA observations of WR 138, with the same format as Fig.~3.}
\end{figure*}

Interferometers measure the fringe contrast or visibility of a source.
The complex visibility of a binary system varies periodically (e.g., Boden et al.~2000):
$$ V = \frac{V_1 + f V_2 \exp{[-2\pi i (u\Delta \alpha+ v\Delta \delta)]}}{(1 + f)} $$
where $\Delta \alpha$ and $\Delta \delta$ represent the binary
separation in right ascension and declination, $u$ and $v$ represent
the spatial frequencies of the Array projected on the sky, $f$
represents the flux ratio of the binary, and $V_1$ and $V_2$ are the
uniform disk visibilities of the primary and secondary components.  In
our case, we assume the stars are unresolved ($V_1 = V_2 = 1.0$), as a
main sequence O star or the emitting region for tau = 1 optical depth
would be smaller than 0.5 mas at the estimated distance of these
systems.  The real and imaginary parts of the complex visibilities are
used to compute the squared visibility amplitude and closure phase to
compare with the observations.

We began our analysis using an adaptive grid search procedure where we
searched for binary solutions through a large grid of separations in
right ascension and declination.  At each step in the grid, we
optimized the position ($\Delta$RA,$\Delta$DEC) and flux ratio of the
binary using the Levenberg-Marquardt least squares minimization
routine mpfit\footnote{http://cow.physics.wisc.edu/\raisebox{0.1em}{\tiny$\sim$\,}craigm/idl/idl.html}
(Markwardt 2009), and computed the $\chi^2$ statistic for each
solution.  We performed the adaptive grid search over a range of
$\pm$16 mas in $\Delta$RA and $\Delta$DEC using 0.1 mas steps and kept
the solution with the lowest $\chi^2$ as the global best-fit binary
model.  If the components in the binary are separated by more than the
coherence length ($\sim$ 8 mas for CLIMB measurements on 250$-$278 m
baselines in the $H$-band), then they will begin to show two separated
fringe packets in the interferometric scans (e.g., Farrington et al.
2010, 2014).  We selected a maximum search range of 16 mas,
corresponding to twice the coherence length, to represent the
separation at which the two fringe packets are completely separated
and no longer overlap.  Beyond this range, binary components of
similar brightnesses would be detected through a visual inspection of
the fringes.  In order to search out to these wider separations, we
added bandwidth smearing to the binary model assuming a rectangular
bandpass profile (Kraus et al. 2005).

This method utilized both measurements of visibility and closure
phase, which helps to remove the 180$^\circ$ ambiguity in the position
angle. The results of these fits are summarized in Table 3, along with
the phasing of the binary systems from Annuk (1990; WR 138) and
Lef\`evre et al. (2005; WR 137), and the fits are shown in the plots of
the data in Figs. 3 and 4. In general, the model works well
for both systems, clearly showing a binary in both cases.  Note that
in the case of a single star, the visibility function follows a simple
Bessel function (e.g., Boyajian et al. 2012), and would also follow a
monotonic decrease in the case of a single star with a large wind
(e.g., P Cygni; Richardson et al. 2013). As the observations appear as
a scatter plot in the upper left panels of Figs. 3 and 4, a resolved
binary is the simplest explanation for the systems.

We examined the resulting $\chi^2$ maps from the grid-searching routine, which show an ambiguity in
the position of the secondary star reflected across the origin.
However, in these mirrored solutions the flux ratio of the binary is
also flipped, so in essence, the solutions are identical.  The
solution with the minimum $\chi^2$ in the maps is reported in Table 3.
There are other possible solutions, however, none
fall within $\Delta \chi^2$ = 3.53 (the 1 $\sigma$ confidence
interval for three fit parameters).  These alternative solutions could
be further ruled out by acquiring additional observations in the
future and fitting the orbital motion directly to the visibilities and
closure phases across multiple epochs.

A visibility near unity would be indicative of a completely unresolved
source. If we are partially resolving the Wolf-Rayet wind or the O
star in these systems, the peak in the visibility curves would be
lower than 1.  For WR 137 and WR 138, the measured visibilities that
are closest to unity are all within the errors from the binary model
that assumes the component stars are unresolved.  Moreover, the number
of data points is not dense enough to include the angular diameters of
the component stars as free parameters in the fit.

\section{Flux Ratios Derived from Spectral Modeling}

According to our interferometric measurements for both WR 137 and WR 138, one component slightly dominates over the other in the H-band.
While calibrations of $H$-band magnitudes with spectral types 
(Martins et al.~2005, Rosslowe \& Crowther 2015) imply that the WR component dominates in the $H$-band in both systems, the typical 
magnitude scatter portrayed by WR stars hinders us to conclude this with certainty.
To ensure that the relative flux contributions are correctly assigned to the correct components of WR 137 and WR 138, 
we performed a spectral analysis of both systems using the Potsdam Wolf-Rayet (PoWR\footnote{PoWR models of Wolf-Rayet stars can be downloaded at
http://www.astro.physik.uni-potsdam.de/PoWR.html}; Hamann \& Gr\"aefner~2004) non-LTE model atmosphere 
code, suitable for any hot stars with winds. Comparing synthetic spectra 
with observations further 
provides us with the fundamental parameters of the components of each system.

A thorough description of the code is beyond the scope of this paper; we only repeat fundamental assumptions here.
The prespecified velocity field $v(r)$ takes the form of the $\beta$-law (Castor et al.~1975)
with $\beta = 1$ and the terminal velocity $v_\infty$ as a free parameter. Clumping is treated 
via the microclumping approach (Hillier 1984). The clumping factor $D$ is treated 
as a free parameter for WR models, and is fixed to $D = 10$ in the winds of O-star models (e.g., Feldmeier et al.~1997). 
The depth-dependent microturbulence parameter $\xi(r)$ grows linearly with $v(r)$ from the photospheric value $\xi_\text{ph}$
to $0.1\,v_\infty$ in the wind, where $\xi_{\rm ph} = 20$\,km s$^{-1}$ for O models and $\xi_{\rm ph} = 50\,$km s$^{-1}$ for WR models (e.g., Shenar et al.~2015).
Macroturbulence is accounted for by convolving the synthetic spectra with appropriate radial-tangential profiles with 
$v_\text{mac} = 30\,$km s$^{-1}$ (e.g., Gray 1975, Bouret et al.~2012).
A more detailed description of the assumptions and methods used
in the code is given by Gr{\"a}efener et al.~(2002), Hamann \& Gr\"afener~(2004), and Sander et al.~(2015).

To analyze the systems, we use a multitude of spectra ranging from the UV to the IR. For both objects, 
all high-resolution, flux calibrated spectra taken with the International Ultraviolet Explorer (IUE) in the spectral 
range $1200 - 2000\,\AA$ were retrieved from the MAST archive. Since the changes with orbital phase in both 
systems are extremely small, we co-added these spectra to obtain a higher signal-to-noise (S/N $\approx 300$) 
for each system. We also retrieved IUE spectra covering the range $1900 - 3100\,\AA$ for both systems 
for a larger coverage of the spectral energy distribution (SED), available in the MAST archive. 
For WR 137, we utilized a high resolution, echelle spectrum from the Keck II telescope and the ESI spectrograph. These data were taken near in time to our interferometry, and covered $\sim 4000$\AA--1$\mu$m, with a typical S/N across the spectrum of 500 or better. 
For WR 138, we use complementary, low resolution ($\Delta \lambda \approx 2\,\AA$) 
optical spectra 
to cover the spectral ranges $3800-4450$ \AA and $6770 - 7400$ \AA, taken between 1991 June 24 -- July 1 with the 2.2m-telescope in Calar Alto, Spain 
(for further details, see Hamann et al.~1995) and low-resolution optical spectra obtained at the Observatoire du Mont M\'egantic (Qu\'ebec) near in time to the interferometric observations, spanning $\sim 4500-6700$\AA.
We also used the NIR spectra obtained for spectral typing of infrared WR stars by Shara et al.~(2012). All spectra were rectified by fitting 
a low-order polynomial to the apparent continuum.

For both systems, we take $UBV$ magnitudes from Neckel et al.~(1980), $R$ and
$I$ magnitude from Monet et al.~(2003), and $JHK$ magnitudes from Kharchenko (2001). The synthetic spectra are convolved with Gaussians of appropriate widths to mimic the spectral resolution. For WR 137, we also recovered MSX photometry from Egan et al.~(2003) and WISE photometry from Cutri et al.~(2012). Fortuitously, these missions observed WR137 close to the 1997 dust-formation maximum and that expected in 2010 (Williams et al. 2001) respectively. The fluxes are marked A, C, D and E (MSX) and W1 and W2 (WISE) in Fig.~6, where they lie well above the stellar model SED and were not used in the fitting

We represent each binary system as a composite of two models corresponding to their component spectra. 
As a first step, we calibrate the fundamental parameters of the components against their spectral types.
These include the effective temperatures $T_*$ and transformed 
radii $R_\text{t}$ (which are related to the mass-loss rates $\dot{M}$, see Schmutz et al.~1989) for the WR stars, 
and $T_*$, $\dot{M}$ and gravities $g_*$ for the O components. We note that $T_*$ and $g_*$ refer to the inner boundary 
of our models, where the mean Rosseland optical depth $\tau_\text{Ross} = 20$. For the O companions, the values of these 
parameters virtually coincide with the photospheric values at $\tau_\text{Ross} = 2/3$.

\begin{figure}
\centering
  \includegraphics[width=0.45\textwidth]{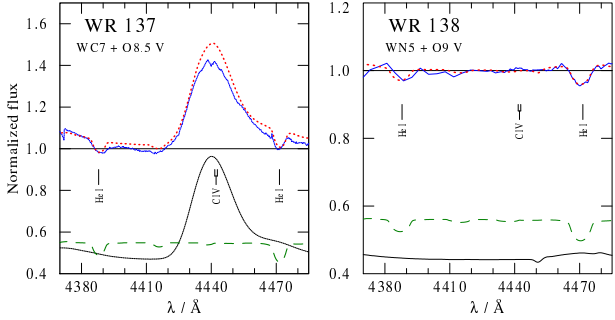}
  \caption{\label{fig:lr} The synthetic composite spectrum (red dotted line) is compared to observations (blue ragged lines) of 
    WR\,137 and 138 in the vicinity of the He\,{\sc i}\,$\lambda 4471$ line. The relative offsets of the WR (black solid line)
    and O (green dashed line) correspond to the relative light contributions in this spectral range. }

\end{figure} 

\renewcommand{\arraystretch}{1.4}

\begin{table}
\normalsize
\footnotesize
\caption{Inferred stellar parameters for WR\,137 and 138} 
\label{tab:stellarpar}
\begin{center}
\tabcolsep=0.1cm
\begin{tabular}{l | c c | c c}
\hline
                                              &                 \multicolumn{2}{c|}{WR 137}         &                           \multicolumn{2}{c}{WR 138}     \\
Component & WR & O & WR & O  \\
% \hline
\hline 
distance $d$ [pc]           & \multicolumn{2}{c|}{$1300$}                &   \multicolumn{2}{c}{$1380$}                    \\
Spectral type               & WC7pd                      & O9\,V           & WN5o              & O9\,V                        \\
$T_{\ast}$ [kK]                               & $60^{+5}_{-5}$                 & $32^{+2}_{-2}$        &  $56^{+5}_{-5}$        & $31^{+2}_{-2}$                   \\
$\log g_*$ [cm\,s$^{-2}$]                     & -                        & $4.0^{+0.3}_{-0.3}$     & -                & $4.0^{+0.3}_{-0.3}$               \\
$\log L$ [$L_\odot$]                          & $5.22^{+0.05}_{-0.05}$            & $4.75^{+0.05}_{-0.05}$   &  $5.35^{+0.05}_{-0.05}$   & $4.82^{+0.05}_{-0.05}$            \\
$\log R_\text{t}$ [$R_\odot$]                 & $0.7^{+0.05}_{-0.05}$             & -               &  $0.9^{+0.05}_{-0.05}$    & -                         \\
$v_{\infty}$ [km s$^{-1}$]                      & $1700^{+100}_{-100}$             & $1800^{+100}_{-100}$    &  $1200^{+100}_{-100}$    & $2000^{+200}_{-200}$                 \\
$R_*$ [$R_\odot$]                             & $3.8^{+1}_{-1}$                    & $7.7^{+1}_{-1}$             &  $4.8^{+2}_{-2}$  & $8.9^{+2}_{-2}$          \\
$D$                       & $4$                       & $10$            &  10              & $10$                       \\
$\log \dot{M}$                                & $-4.65^{+0.2}_{-0.2}$       & $-7.1^{+1.0}_{-0.3}$  &  $-5.15^{+0.2}_{-0.2}$   &  $-7.7^{+0.5}_{-1.5}$     \\
$v \sin i$ [km s$^{-1}$]                             & -                        & $220^{+20}_{-20}$      &  -               & $350^{+30}_{-30}$                      \\
$M_{V}$ [mag]                     & $-4.18^{+0.2}_{-0.2}$                  & $-4.34^{+0.2}_{-0.2}$         &  $-4.40^{+0.2}_{-0.2}$         & $-4.53^{+0.2}_{-0.2}$        \\
$M_{H}$ [mag]                     & $-4.07^{+0.2}_{-0.2}$                  & $-3.41^{+0.2}_{-0.2}$         &  $-4.28^{+0.3}_{-0.3}$         & $-3.60^{+0.3}_{-0.3}$         \\
$E_{B-V}$ [mag]                             &  \multicolumn{2}{c|}{$0.74^{+0.02}_{-0.02}$} &  \multicolumn{2}{c}{$0.67^{+0.02}_{-0.02}$}                       \\
$A_V$ [mag]                                 & \multicolumn{2}{c|}{$2.29^{+0.06}_{-0.06}$}  & \multicolumn{2}{c}{$2.08^{+0.06}_{-0.06}$}                                       \\
\hline
\end{tabular}
\end{center}
\end{table}

With the parameters fixed, the light ratio can be derived by 
examining a multitude of features in the available spectra, primarily in the UV and optical. 
An example is shown in Fig.\,\ref{fig:lr}, where the best-fitting models for both systems are compared to optical observations in the vicinity 
of the He\,{\sc i}\,$\lambda 4471$ line. This line maintains an almost identical equivalent width 
in the temperature domain $30 - 35\,$kK, which is the relevant temperature domain for both O companions concerned here.
Its strength is therefore directly related to the dilution factor, i.e., the light ratio in the visual.
In contrast, the
He\,{\sc i}\,$\lambda 4388$ line, also shown in Fig.\,\ref{fig:lr}, is sensitive to $T_*$ and $g_*$. The derived 
light ratios are consistent with a multitude of features in the available spectra. We further confirmed 
our results by comparing with equivalent width calibrations for single O stars (e.g.,Conti \& Alschuler~1971).

\begin{figure*}
\centering
  \includegraphics[width=0.9\textwidth]{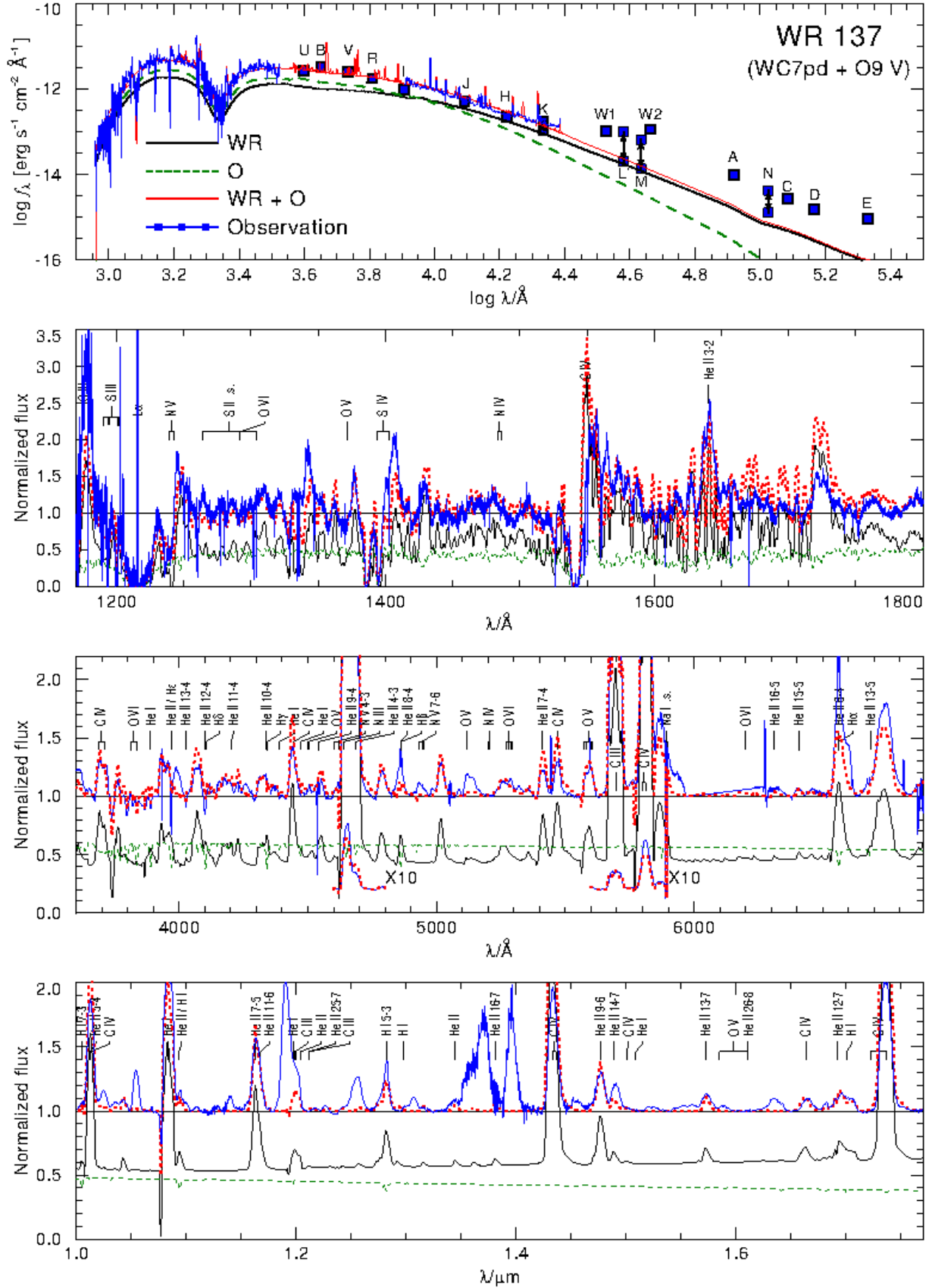}
  \caption{
  The best fitting synthetic SED (upper panel) and normalized spectrum (lower panels) compared to observations of WR\,137 (blue lines and squares). The composite model
   is the superposition of the WR (black solid line) and O (green dashed line) models. The relative offsets of the model continua shown in the lower panels
   depict the relative light contribution in each spectral range. The few arrows shown in the upper panel illustrate the IR excess variability in the system from Williams et al.~(2001). Note that the CHARA observations were collected in the $H$-band ($\log \lambda \approx 4.2$).
}
\label{fig:WR137SED}
\end{figure*}

\begin{figure*}
\centering
  \includegraphics[width=0.9\textwidth]{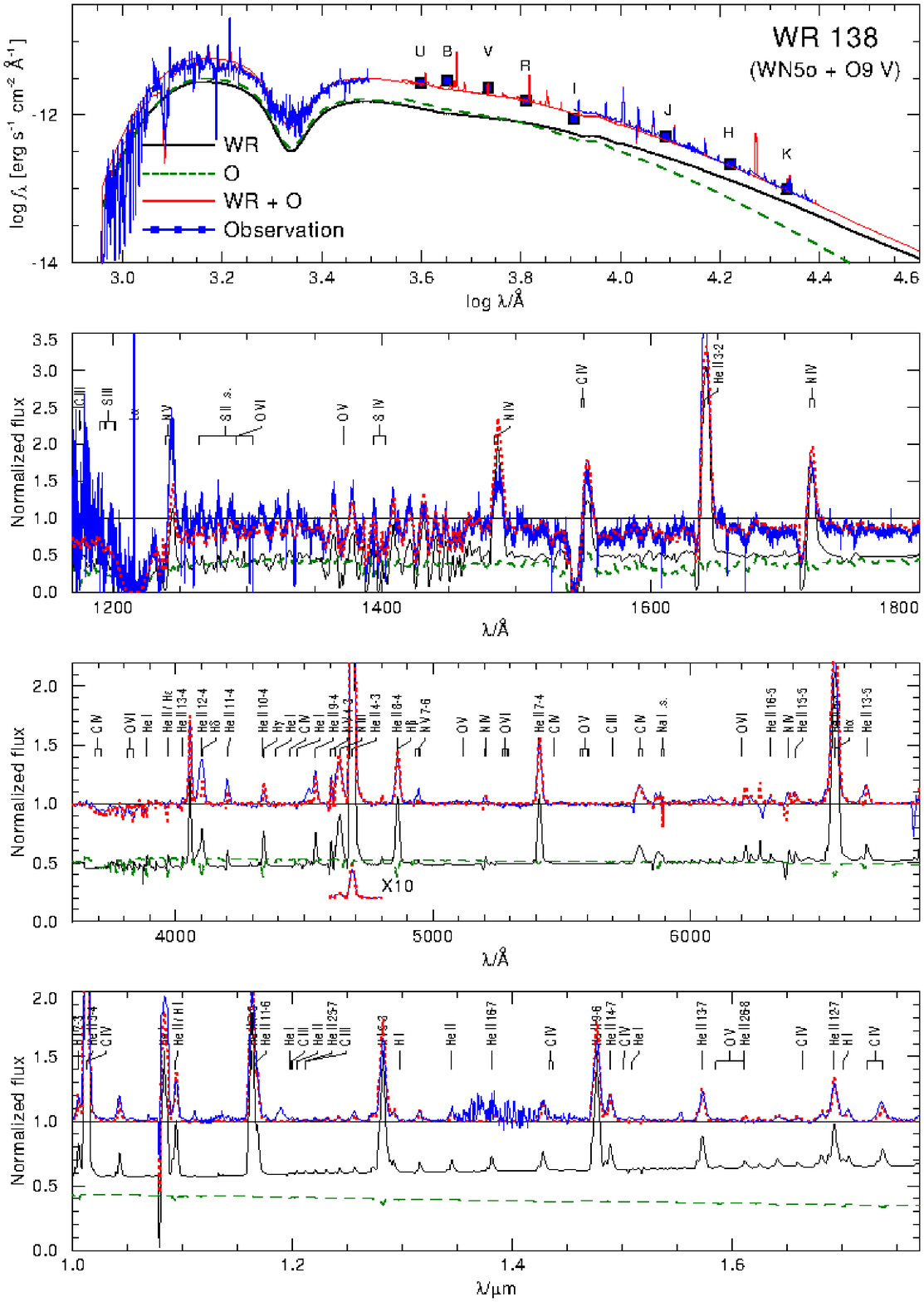}
  \caption{As Fig.\,\ref{fig:WR137SED}, but for WR\,138. Note that the CHARA observations were collected in the $H$-band ($\log \lambda \approx 4.2$)}
\label{fig:WR138SED}
\end{figure*}

The stellar parameters are then further refined until no significant improvement can be achieved. Wind parameters 
are derived based on the strengths and shapes of emission lines. For the O stars, $v_\infty$ could be constrained 
from the C\,{\sc iv} and Si\,{\sc iv} resonance lines. Their mass-loss rates are very roughly constrained based on the existence
of P Cygni or emission features clearly associated with the O components or lack thereof. The projected rotational 
velocity $v \sin i$ is derived primarily from He\,{\sc i} and He\,{\sc ii} lines (see e.g.\ Fig.\,\ref{fig:lr}).
The components' luminosities $L$ 
and the reddening $E_{B-V}$ are derived by comparing the composite synthetic spectrum to the spectral energy 
distribution (SED) of each system. We assume the reddening law given by Seaton (1979) with $R_V = 3.1$. 
We adopt distances from Rosslowe \& Crowther (2015), which result in consistent luminosities for both parameters.

The SEDs and spectra along with the best-fitting models are shown in Figs.~\ref{fig:WR137SED} 
and \ref{fig:WR138SED}. The derived parameters are listed in Table\,\ref{tab:stellarpar}. Uncertainties are estimated based on the sensitivity 
of the fit quality to changes in the corresponding parameter. The Table also gives stellar radii $R_*$ 
(calculated via the Stefan-Boltzmann law), $V-$ and $H-$band magnitudes, and extinctions $A_\text{V}$. Our results turn out to be very consistent with what is expected for stars of the given spectral types. The derived light ratios 
indeed imply that the WR component is the dominant source in the $H$-band in both systems.  
For WR 137, the WR component contributes (59$\pm$4)\% of the $H$-band flux as derived by the interferometry, and is seen to contribute a fractional flux of 0.59 in the PoWR model.
Similarly, for WR 138, we measured the WR component to contribute (67$\pm$1)\% of the $H$-band flux, with the PoWR results show the fraction to be 0.66. These measurements show the strong agreement in the two methods, and strengthen the results of both investigations.

\section{Discussion}

The interferometric measurements of WR 138 represent the first time a WN star has been resolved in a spectroscopic binary. We note that the fundamental parameters derived through our spectroscopic modeling of the binary system agree with the expectations for a late O dwarf (Martins et al.~2005, Martins \& Plez 2006) as the companion. The absorption lines in the system are very broad and shallow (see Figs.~5 and 7), and have caused some controversy in the literature. The star was first called a binary based upon the absorption lines of hydrogen and neutral helium in the blue by Hiltner (1945). Massey (1980) studied a time-series of moderate-resolution photographic spectra to find that the semi-amplitude of any binary would likely be $\lesssim 30$ km s$^{-1}$ for periods less than 6 months, for which his radial velocity study is most sensitive. 

Lamontagne et al.~(1982) examined a longer time-series of spectra of WR 138, including the measurements of Massey (1980) and Bracher (1966). They found a long-period orbit, with a period of $\sim1700$ d, and attributed short-term variability of the WR emission lines as arising from a short-period, 2.3 d, orbit with a neutron star. The short-period orbit was not confirmed when Annuk (1990) later examined the system and concluded that the system was a spectroscopic binary, with a period of $\sim 1500$ d. The O star is fairly exotic compared to most O dwarfs, as the measured value of $v \sin i$ is roughly $500$ km s$^{-1}$ (Hamann et al. 2006, Massey 1980, Annuk 1990), while our spectral modeling indicates a lower value of $\sim 350$ km s$^{-1}$. This is incredibly high in comparison to the population of O stars both in the Galaxy (Howarth et al.~1997), or even in the massive, star-forming region of 30 Dor (Ram\'irez-Agudelo et al.~2013), where the measured $v \sin i$ would place the companion in the top $5-10$\% of the population. 

The period derived by Annuk (1990) and Palate et al.~(2013) is quite long (4.2 yr), and should be longer than typical orbits associated with binary interactions (e.g., Sana et al.~2012). However, the period limit for a zero-age main-sequence O star to experience spin-up from a companion from the results of Sana et al.~(2012) is actually similar to the orbital period of WR 138. Therefore, the extreme rapid rotation of this companion may actually be the remnant of a past interaction where the current WR star lost mass via Roche lobe overflow which then deposited angular momentum and mass onto the O star. If the semi-major axis is conserved in this process, then if the original system were $30+20 M_\odot$ stars that interacted to become $15+25 M_\odot$ stars through a combination of mass loss and accretion, the period would have originally been shorter ($< 1400$ d) and near the adopted upper limits for an interacting binary as defined by Sana et al.~(2012).

These interferometric observations of WR 137 represent the second reported binary measurements of this star. Rajogopal (2010) found a binary flux ratio equal to our measurements. 
However, the projected separations between our measurements and those of Rajagopal (2010) differ by about a factor of two, and they differ by $\sim180^\circ$, but were taken at opposite quadratures. Two measurements of separation and position angle from two different instruments are not enough to warrant an attempt of a visual orbit. Further, we note that we are unsure if Rajagopal (2010) accounted for the contamination of emission lines in the interferometry, which can alter the measured separation and position angle. However, it seems that the eventual orbital solution will favor a system with a nearly edge-on geometry. 

A directly edge-on system will have masses that are perhaps lower than expected in the WR 137 system, but there are supporting reasons to trust the inclination to be closer to edge-on. Marchenko et al.~(1999) imaged the system in the $H$- and $K$-bands with the {\it HST} and the NICMOS camera. These images were taken during and after a periastron passage when dust is formed. These images revealed structures in the $K$-band, but not in the $H$-band, that were thought to have emerged from the wind-wind collision region. It is interesting that the position angle of the structure imaged by {\it HST} is $\approx 110^\circ$. If the positions measured by the IOTA and CHARA interferometers are indeed pointing towards an edge-on orbit, then the dust formation may occur near the orbital plane. Further observations can elucidate the details of the orbit and the formation region of the dust.
We also note that the opening angle of the shock cone is $\approx 20^\circ$ according to the approximations given by Gayley (2009) and our spectral modeling, so the small opening angle in the {\it HST}/NICMOS imaging is consistent with this explanation.

Clearly, the next step in these systems is two-fold. We need to resolve the orbits at multiple epochs in order to measure the visible orbit of the systems. This will lead to constraints on the inclinations of the system and the orbit. Contemporaneous to this, we need to re-determine the double-lined spectroscopic orbits of these systems as both orbits are poorly constrained for the O star components. The O star in the WR 138 system is extremely difficult to measure due to its large value for $v \sin i$, so high resolution spectroscopy with high S/N is needed. The resolution and interferometric followup of WR 137 and WR 138 will double the number of the masses measured for Wolf Rayet stars in longer period orbits, with WR 138 being the first WN star with a visual orbit. Future interferometry of WR 137 would benefit from measurements made in both the $H-$ and $K-$band near in time so that the the binary parameters can first be resolved in the $H-$band. These derived parameters can then be used to pinpoint the exact place where the dust formation happens in the close environs of the system. Such observations can be used to further understand the $K$-band {\it HST} imaging of the system that was reported by Marchenko et al.~(1999). WR 137 provides an exciting example of a WC$+$O system where we can determine empirically the location of the dust formation through long-baseline infrared interferometry and then compare the close environs of the system's dust production to the larger scale imaging reported by Marchenko et al.~(1999).

\section*{Acknowledgements}

We wish to thank Michael Shara, Jacqueline Faherty, and Graham Kanarek for allowing us to use their NIR spectra of these two stars. We thank the CHARA staff for supporting these observations and the Mount Wilson Institute for its continued support of the CHARA Array. This material is based upon work supported by the National Science Foundation
under Grants AST-1211929 and AST-1411654. Some of the spectroscopic data were obtained at the W. M. Keck Observatory on the summit of Mauna Kea. The authors wish to recognize and acknowledge the very significant cultural role and reverence that the summit of Mauna Kea has always had within the indigenous Hawaiian community. We are most fortunate to have the opportunity to conduct observations from this mountain. The {\it IUE} data presented in this paper were obtained from the Mikulski Archive for Space Telescopes (MAST). STScI is operated by the Association of Universities for Research in Astronomy, Inc., under NASA contract NAS5-26555. Support for MAST for non-HST data is provided by the NASA Office of Space Science via grant NNX09AF08G and by other grants and contracts.

NDR acknowledges postdoctoral support by the University of Toledo and by the Helen Luedtke Brooks Endowed Professorship, and is thankful for his former CRAQ (Qu\'ebec) fellowship which supported him at the beginning of this research. TS is grateful for financial support from the Leibniz Graduate School for Quantitative Spectroscopy in Astrophysics, a joint project of the Leibniz Institute for Astrophysics Potsdam (AIP) and the institute of Physics and Astronomy of the University of Potsdam. AFJM and NSL are grateful for financial aid from NSERC (Canada) and FQRNT (Quebec). PMW is grateful to the Institute for Astronomy for continued hospitality and access to the facilities of the Royal Observatory Edinburgh.

\bsp \label{lastpage}


\begin{thebibliography}{99}
%
\bibitem[Aldoretta et al.(2015)]{emily}
	Aldoretta, E. J., Caballero-Nieves, S. M., Gies, D. R., et al.
	2015, AJ, 149, 26
\bibitem[Annuk(1990)]{annuk}
	Annuk, K. 1990, AcA, 40, 267
\bibitem[Barbier (1963)]{barbier}
	Barbier, M.
	1963, POHP, 6, 36
\bibitem[Boden (2000)]{boden}
	{Boden}, A.~F. 2000, in Principles of Long Baseline Stellar
	Interferometry, ed. P.~R. {Lawson} (Pasadena, CA: JPL), 9
\bibitem[Bouret et al.(2012)]{bouret}
	Bouret, J. C., Hillier, D. J., Lanz, T., \& Fullerton, A. W.
	2012, A\&A, 544, 67
\bibitem[Boyajian et al.(2012)]{tabby12}	
	Boyajian, T. S., McAlister, H. A., van Belle, G., et al.
	2012, ApJ, 746, 101
\bibitem[Boyajian et al.(2014)]{tabby14}	
	Boyajian, T. S., van Belle, G., \& von Braun, K.
	2014, AJ, 147, 47
\bibitem[Bracher(1966)]{bracher}
	Bracher, K.
	1966, Ph.D. thesis, Indiana University
\bibitem[Castor et al.(1975)]{cak}
	Castor, J. I., Abbott, D. C., \& Klein, R. I.
	1975, ApJ, 195, 157
\bibitem[Conti \& Alschuler(1971)]{conti1971}
	Conti, P. S., \& Alschuler, W. R.
	1971, ApJ, 170, 325
\bibitem[Cutri et al.(2012)]{cutri}
	Cutri, R. M., et al. 
	2012, VizieR Online Data Catalogs: II/311	
\bibitem[de Marco \& Schmutz(1999)]{deM99}
	De Marco, O., \& Schmutz, W.
	1999, A\&A, 345, 163
\bibitem[de Marco et al.(2000)]{deM00}
	De Marco, O., Schmutz, W., Crowther, P. A., et al.
	2000, A\&A, 358, 187	
\bibitem[Egan et al.(2003)]{egan}
	Egan, M. P., Price, S. D., Kraemer, K. E., et al.
	2003, VizieR Online Data Catalogs: V/114
\bibitem[Fahed et al.(2011)]{fahed}
	Fahed, R., Moffat, A. F. J., Zorec, J., et al.
	2011, MNRAS, 418, 2
\bibitem[Farrington et al.(2010)]{chris10}
	Farrington, C. D., ten Brummelaar, T. A., Mason, B. D., et al.
	 2010, AJ, 139, 2308
\bibitem[Farrington et al.(2014)]{chris}
	Farrington, C. D., ten Brummelaar, T. A., Mason, B. D., et al.
	2014, AJ, 148, 48
\bibitem[Feldmeier et al.(1997)]{feldmeier}
	Feldmeier, A., Puls, J., \& Pauldrach, A. W. A.
	1997, A\&A, 322, 878
\bibitem[Gayley (2009)]{gayley}
	Gayley, K. G.
	2009, ApJ, 703, 89
\bibitem[Gr\"aefener et al.(2002)]{Graefener2002}
	Gr{\"a}fener, G., Koesterke, L., \& Hamann, W.-R.
	2002, A\&A, 387, 244
\bibitem[Gray (1975)]{gray}
	Gray, D. F.
	1975, ApJ, 202, 148
\bibitem[Hamann \& Gr\"aefner(2004)]{hamann2004}
	Hamann, W.-R., \& Gr\"aefener, G.
	2004, A\&A, 427, 697
\bibitem[Hamann et al.(2006)]{hamann2006}
	Hamann, W.-R., Gr\"aefener, G., \& Liermann, A.
	2006, A\&A, 457, 1015
\bibitem[Hamann et al.(1995)]{hamann95}
	Hamann, W.-R., Koesterke, L., \& Wessolowski, U.
	1995, Astron. \& Astrophys. Suppl., 113, 459.
\bibitem[Hanbury Brown et al.(1970)]{hanbro70}
	Hanbury Brown R., Davis J., Herbison-Evans D., Allen L. R.
	1970, MNRAS, 148, 103
\bibitem[Hillier(1984)]{hillier84}
	Hillier, D. J.
	1984, ApJ, 280, 744
\bibitem[Hiltner(1945)]{hiltner}
	Hiltner, W. A.
	1945, ApJ, 101, 356
\bibitem[HONEYCUTT R.K.; McCUSKEY S.W.]{honeycuttham}
	Honeycutt, R. K., \& McCuskey, S. W.
	1966, PASP, 78, 289	
\bibitem[Howarth et al.(1997)]{howarth}
	Howarth, I. D., Siebert, K. W., Hussain, G. A. J., Prinja, R. K.
	1997, MNRAS, 284, 265
\bibitem[Kharchenko(2001)]{Kharchenko2001}
	Kharchenko, N. V.
	2001, KFNT, 17, 409
\bibitem[Kraus et al.(2005)]{kraus}
	{Kraus}, S., {Schloerb}, F.~P., {Traub}, W.~A., {et~al.} 
	2005, AJ, 130, 246
\bibitem[Lafrasse et al.(2010)]{lafrasse}
	Lafrasse S., Mella G., Bonneau D., et al.
	2010, Proc. SPIE, 7734, 140
\bibitem[Lamontagne et al.(1982)]{robert}
	Lamontagne, R., Koenigsberger, G., Seggewiss, W., \& Moffat, A. F. J.
	1982, AJ, 253, 230 
\bibitem[Lefevre(2005)]{lefevre}
	Lef\`evre, L., Marchenko, S. V., L\'epine, S., et al.
	2005, MNRAS, 360, 141
\bibitem[Ljunggren \& Oja (1961)]{swedes}
	Ljunggren, B., \& Oja, T.
	1961, Uppsala Astron. Obs. Ann., 4, 1 
\bibitem[Marchenko(1999)]{dusty}
	Marchenko, S. V., Moffat, A. F. J., \& Grosdidier, Y.
	1999, ApJ, 522, 433
\bibitem[Markwardt (2009)]{markwardt}
	Markwardt, C. B. 2009, in Astronomical Society of the Pacific Conference Series, Vol. 411, 
	Astronomical Data Analysis Software and Systems XVIII, ed. D. A. Bohlender, D. Durand, \& P. Dowler, 251
\bibitem[Martins \& Plez(2006)]{mp06}
	Martins, F., \& Plez, B.
	2006, A\&A, 457, 637
\bibitem[Martins et al.(2005)]{martins}
	Martins, F., Schaerer, D., \& Hillier, D. J.		
	2005, A\&A, 436, 1049
\bibitem[Massey(1980)]{massey}
	Massey, P. 1980, ApJ, 236, 526
\bibitem[Millour et al.(2009)]{vltigamvel}
	Millour F., Driebe T., Chesneau O., et al.
	2009, A\&A, 506, L49
\bibitem[Millour et al.(2007)]{vltigamvel}
	Millour F., Petrov, R. G., Chesneau O., et al.
	2007, A\&A, 464, 107	
\bibitem[Monet et al.(2003)]{monet}
	Monet, D. G., Levine, S. E., Canzian, B., et al.
	2003, AJ, 125, 984
\bibitem[Monnier et al.(2004)]{monnier04}
	Monnier, J.~D., Traub W. A., Schloerb F. P., et al.
	2004, ApJ, 602, L57
\bibitem[Monnier et al.(2011)]{monnier11}
	Monnier, J.~D., Zhao, M., Pedretti, E., et al.
	2011, ApJ, 742, L1
\bibitem[Neckel et al.(1980)]{Neckel1980}
	Neckel, T., Klare, G., \& Sarcander, M.
	1980, BICDS, 19, 61
\bibitem[North et al.(2007)]{north}
	North, J.~R., Tuthill P. G., Tango W. J., \& Davis J.
	2007, MNRAS, 377, 415
\bibitem[Nujis]{nujis}
	Nugis, T. \& Lamers, H. J. G. L. M. 
	2000, A\&A, 360, 227
\bibitem[Palate et al.(2013)]{palate}
	Palate, M., Rauw, G., De Becker, M., Naz\'e, Y., \& Eenens, P.
	2013, A\&A, 560, 27
\bibitem[Pauls et al.(2005)]{pauls05} 
	Pauls, T. A., Young, J. S., Cotton, W. D., \& Monnier, J. D. 
	2005, PASP, 117, 1255
\bibitem[Rajagopal(2010)]{iota}
	Rajagopal, J.
	2010, in Revista Mexicana de Astronomia y Astrofisica Serie de Conferencias, Vol. 38, 54
\bibitem[Ramirez et al.(2013)]{ram13}
	Ram\'irez-Agudelo, O. H., Sim\'on-D\'iaz, S., Sana, H., et al.
	2013, A\&A, 560, 29	
\bibitem[\protect\citeauthoryear{Richardson et al.}{2013}]{richardson2013} 
	Richardson, N. D., Schaefer, G. H., Gies, D. R., et al. 
	2013, ApJ, 769, 118
\bibitem[Rosslowe \& Crowther(2015)]{rc15}
	Rosslowe, C. K., \& Crowther, P. A.
	2015, MNRAS, 447, 2322
\bibitem[Sana et al.(2014)]{sana14}
	Sana, H., de Mink, S., de Koter, A., et al.
	2012, Science, 337, 444
\bibitem[Sana et al.(2014)]{sana14}
	Sana, H., Le Bouquin, J.-B., Lacour, S., et al.
	2014, ApJS, 215, 15
\bibitem[Sander et al.(2015)]{sander}
	Sander, A., Shenar, T., Hainich, R., et al.
	2015, A\&A, 577, 13
\bibitem[Schaefer et al.(2014)]{sch14} 
	Schaefer, G. H., ten Brummelaar, T., Gies, D. R., et al. 
	2014, Nature, 515, 234
\bibitem[Schmutz et al.(1989)]{schmutz89}
	Schmutz, W., Hamann, W.-R., \& Wessolowski, U.
	1989, A\&A, 210, 236
\bibitem[Schmutz et al.(1997)]{schmutz97}
	Schmutz, W., Schweickhardt, J., Stahl, O., et al.
	1997, A\&A, 328, 219
\bibitem[Seaton (1979)]{seaton1979}
	Seaton, M. J.
	1979, MNRAS, 187, 73
\bibitem[Shenar et al.(2015)]{tomer}
	Shenar, T., Oskinova, L., Hamann, W.-R., et al.
	2015, ApJ, 809, 135
\bibitem[Shara et al.(2012)]{shara}
	Shara, M. M., Faherty, J. K., Zurek, D., et al.
	2012, AJ, 143, 149
\bibitem[Smith et al.(1996)]{smith96}
	Smith, L. F., Shara, M. M., \& Moffat, A. F. J. 1996, MNRAS, 281, 163
\bibitem[Smith \& Owocki(2006)]{smith}
	Smith, N., \& Owocki, S. 
	2006, ApJ, 645, L45
\bibitem[{ten Brummelaar et al.}{2005}]{charainst}
	ten Brummelaar, T.~A., McAlister, H.~A., Ridgway, S.~T., et al. 
	2005, ApJ, 628, 453
\bibitem[{ten Brummelaar et al.}{2013}]{climb} 
	ten Brummelaar, T.~A., Sturmann, J., Ridgway, S.~T., et al. 
	2013, JAI, 2, 1340004 
\bibitem[Tuthill  et al.(2008)]{tuthill}
	Tuthill, P., Monnier, J. D., Lawrance, N., et al.
	2008, ApJ, 675, 698
\bibitem[Williams et al.(2001)]{williams}
	Williams, P. M., Kidger, M. R., van der Hucht, K. A., et al.
	2001, MNRAS, 324, 156
\bibitem[Williams et al.(1985)]{williams85}
	Williams, P. M., Longmore, A. J., van der Hucht, K. A., et al.
	1985, MNRAS, 215, 23
\bibitem[Williams et al.(2009)]{williams09}
	Williams, P. M., Marchenko, S. V., Marston, A. P.,  et al.
	2009, MNRAS, 395, 1749	




\end{thebibliography}
\end{document}